\documentclass[journal]{vgtc}                     

\onlineid{0}

\vgtccategory{Research}

\title{Visual Cue Interactions in AR-Guided Needle Insertion: A Prostate Biopsy-Inspired Phantom Study}

\author{Xinrui Zou,
\and Mingxu Liu,
\and Thomas T. Jones,\\ %
\and Braden Millan,\\ %
\and Sandeep Gurram,\\ %
\and Peter A. Pinto,\\ %
\and Raisa Z. Freidlin,\\ %
\and \authororcid{Alejandro Martin-Gomez}{0000-0001-9341-3477}\\ %
}

\authorfooter{
  \item
  	Xinrui Zou is with Johns Hopkins University and National Institute of Biomedical Imaging and Bioengineering, NIH. 
    E-mail: xzou8@jhu.edu
  \item
  	Mingxu Liu is with Johns Hopkins University. 
  \item Thomas T. Jones and Raisa Freidlin are with National Institute of Biomedical Imaging and Bioengineering, NIH. 
  \item Braden Millan, Sandeep Gurram and Peter A. Pinto are with Urologic Oncology Branch, National Cancer Institute, NIH. 
  \item Alejandro Martin-Gomez is with University of Arkansas. 
  Email: alejandro.martin@uark.edu
}

\abstract{
Despite the apparent simplicity of the motor action involved during percutaneous needle procedures, manipulating the tool's direction becomes challenging when clinicians cannot directly visualize internal anatomy and must rely on ultrasound images, which increase cognitive demand.
 Augmented reality (AR) offers the promise to assist with these tasks by providing pertinent visual information in the clinician’s field of view.
 However, simply providing visual information that ignores meaningful visual cues can complicate depth perception and spatial understanding.
 \noindent
 In this work, we introduce and evaluate three visualization techniques for needle alignment developed during design sessions with medical experts: a localized \textit{focus-and-context window}, a \textit{color-based proximity encoding}, and an \textit{explicit trajectory overlay}.
These techniques were evaluated in a user study (n=26) including clinical experts (n=7) using a prostate-biopsy-inspired phantom.
Results from this study suggest that cue effects depended on the surrounding cue configuration and user expertise. For novices, \textit{explicit trajectory overlay} improved targeting accuracy and reduced retreat behavior, but its effect on completion time varied across cue configurations, with slower performance when the overlay was presented alone. For experts, the \textit{focus-and-context window} reduced completion time and retreat events, while \textit{color-based proximity overlay} improved completion time. Subjectively, color cues were often perceived as helpful even when their effects on accuracy were not consistent. These results suggest that AR guidance strategies for percutaneous interventions should consider user expertise and visual context.} 


\teaser{
  \centering
  \includegraphics[width=\linewidth]{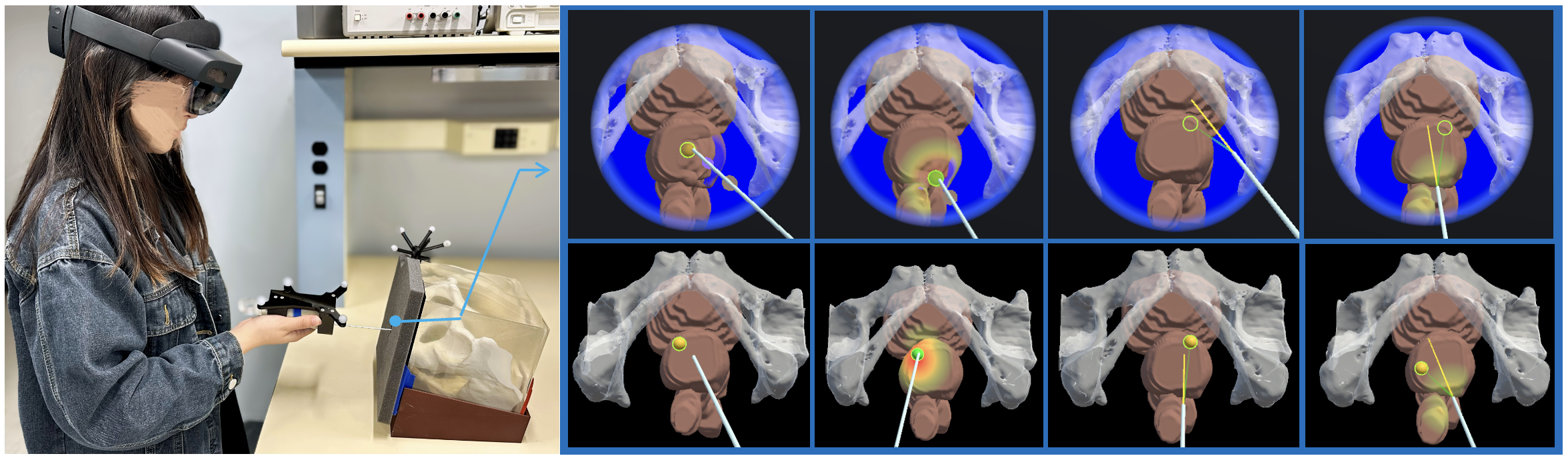}
  \caption{System overview and visualization combinations. Left: A participant using HoloLens 2 with a tracked biopsy needle and a gel-embedded pelvis phantom. Right: Overlays rendered from the user’s perspective, showing all $2{\times}2{\times}2$ combinations of the three visualization cues: Focus-and-context Window (W), Color-proximity (C), and Trajectory guidance (T). Top row (W on): W, W+C, W+T, W+C+T. Bottom row (W off): None (baseline), C, T, C+T. Additional details are provided in ~\cref{sec:Methods}.}
  \label{fig:teaser}
}

\graphicspath{{figs/}{figures/}{pictures/}{images/}{./}} 

\usepackage{booktabs}                  
\usepackage{lipsum}                    
\usepackage{mwe}                       
\usepackage{ccicons}                   

\usepackage{mathptmx}                  
\usepackage{amsmath}                   
\usepackage{graphicx}
\usepackage{subcaption}

\begin{document}


\firstsection{Introduction}

\maketitle

Percutaneous needle procedures are spatially demanding interventions that require clinicians to navigate small instruments through soft tissue toward deep or narrow targets, often with limited visual or tactile feedback. These procedures are common in clinical practice, ranging from biopsies and drainage to localized injections and tumor ablations. Among these procedures, prostate biopsy provides a representative and clinically relevant context because accurate and efficient tissue sampling requires significant experience.
Inaccurate insertion can lead to false negatives or suboptimal sampling, delaying diagnosis and compromising downstream treatment decisions \cite{ahmed2017diagnostic}. These clinical risks motivate improved spatial guidance.

To guide such procedures, clinicians rely on medical imaging. Computed Tomography (CT) and Magnetic Resonance Imaging (MRI) are commonly used for pre-operative planning due to their high spatial resolution and soft tissue contrast, respectively. Ultrasound (US), in contrast, is widely used for intraoperative guidance because it provides real-time imaging at the bedside. Across modalities, images are usually shown on external monitors rather than in direct alignment with the patient’s body, forcing clinicians to mentally register 2D slices with the 3D physical anatomy during the intervention \cite{evans2025augmented}. The spatial mismatch between the patient's body and the imaging display introduces several challenges: disrupted hand–eye coordination, increased cognitive load, and reduced spatial awareness, especially when navigating toward deep, small, or shifting targets.

Augmented reality (AR) has emerged as a promising approach to mitigate these challenges by superimposing medical imaging content, such as anatomical structures, tool trajectories, or internal targets, directly onto the patient's body. Prior studies have shown that AR can improve targeting accuracy and reduce task completion times during the performance of percutaneous needle procedures \cite{heinrich2020comparison, park2020augmented}. This technology offers valuable information for spatial guidance. However, its effectiveness depends not only on \textit{what} but also on \textit{how} the information is presented. A central challenge in achieving intuitive and accurate spatial understanding lies in designing overlays that convey depth and spatial relationships clearly, without introducing visual clutter or ambiguity. For example, overlays lacking strong perceptual cues, such as occlusion or shading, can appear disconnected from anatomy, reducing the operator’s ability to judge distances and positions accurately \cite{gsaxner2021augmented}.

To support spatial understanding in AR-guided interventions, researchers have explored a range of visualization strategies that differ in both their visual structure and the types of depth cues they provide. 

\textit{Focus-and-context} techniques enable the visualization of internal anatomical structures through virtual cutaways or synthetic windows \cite{fuchs1996towards}. By leveraging occlusion, the strongest depth cue known, these methods reveal anatomical targets by obscuring the surrounding anatomy. At the same time, they provide minimal but valuable information regarding the occluding structures around the targets, supporting enhanced spatial judgment. Because of these characteristics, focus-and-context approaches help users interpret layered anatomical relationships during tool navigation.

Another family of techniques relies on \textit{illustrative encoding}, where artificial visual mappings serve to convey depth or proximity. A classic example of these approaches is chromostereopsis, a rendering technique that assigns warmer colors to near objects and cooler tones to distant ones, creating a perceptual gradient that helps distinguish spatial layers~\cite{kersten2013evaluation}. Unlike occlusion or shading, these cues require user interpretation and are intentionally designed to enhance depth understanding without modifying object geometry.

Explicit \textit{trajectory-based guidance} has often been used in needle-based procedures to improve tool alignment. This includes visualizing the optimal insertion path as a color-coded trajectory, indicating entry points, or dynamically updating indicators to reflect deviation from the ideal route \cite{mewes2018concepts}. Such methods offer direct feedback during interaction by encoding how well the user’s actions align with the intended task.

Collectively, these strategies address complementary aspects of spatial understanding, yet existing works often evaluate them in isolation. Combining multiple cues is not guaranteed to yield additive benefits because overlays may compete for attention, increase visual clutter, or cause unfavorable speed–accuracy trade-offs. 

To address the limited understanding of how AR visual cues interact across user expertise levels, we conducted the same controlled factorial protocol in two cohorts: novices (Study I) and clinical experts (Study II). This design allowed us to analyze overall performance trends across all participants and test whether cue effects varied by expertise. We also conducted within-cohort analyses to characterize expertise-specific effects.
We systematically evaluated three representative techniques: \textit{focus-and-context}, \textit{color-based illustrative}, and \textit{explicit trajectory guidance}. These techniques were implemented in a unified AR platform to support controlled execution and assessment of needle-alignment tasks, with prostate biopsy serving as a representative procedural scenario. Accordingly, our evaluation focuses on task performance and user perception in a phantom-based alignment task, rather than clinical efficacy.
By evaluating accuracy, task duration, and workload, our contributions are threefold: (1) a unified implementation of three representative AR visualization strategies for needle-alignment guidance;
(2) a factorial, multi-metric evaluation of their main and interaction effects, including pooled and expertise-specific analyses; and (3) empirical evidence suggesting that cue benefits are expertise-dependent and not uniformly additive, yielding design considerations for AR-guided needle interventions.

\section{Related Work}

Augmented reality has been used for image-guided needle interventions for over twenty years, with studies consistently showing improved targeting accuracy when virtual overlays are presented in situ. Randomized trials using phantoms have demonstrated that providing guidance using AR head-mounted displays significantly reduces targeting error compared to the standard ultrasound method during breast biopsies \cite{rosenthal2002augmented}. Subsequent studies have observed similar benefits in ultrasound-guided phantom biopsies, where AR reduced mean insertion error without increasing task duration \cite{ruger2020ultrasound}. In central line placement procedures, AR navigation systems have also increased success rates for both novice and experienced clinicians \cite{li2025align}. These improvements are largely attributed to AR’s ability to reduce cognitive workload by integrating visual guidance directly into the procedural field, preserving hand-eye coordination and supporting spatial judgment during tool alignment \cite{heinrich2020comparison}.

Although AR guidance can contribute to a reduction in the cognitive load, an overabundance of visual cues can conversely strain the user. Existing studies involving expert interventionists have shown that AR systems that present cluttered or distracting information can lead to subtle benefits and increased mental workload during the performance of routine tasks \cite{li2025align}. These findings emphasize the relevance of properly presenting visual information to the user. Effective AR designs should balance information richness with clarity, ensuring that virtual overlays enhance, rather than disrupt, the practitioner’s situational understanding. This has motivated exploration of various visualization strategies to convey depth and guidance cues in an intuitive yet non-overwhelming manner.

Various AR visualization techniques have been proposed to improve depth perception and spatial guidance in image-guided interventions, particularly when clinicians must interpret anatomy through indirect, incomplete, or visually constrained views \cite{bajura1992merging, kalkofen2007interactive, zou2024arthronerf}. One approach involves the use of focus-and-context overlays, which aim to reveal internal structures while preserving surrounding anatomical context. This class of techniques includes virtual cutaways \cite{bajura1992merging} and structured occlusion \cite{kalkofen2007interactive}, all of which leverage occlusion cues to help users perceive depth and structure. Early AR systems demonstrated that such visualizations can enhance the user’s ability to judge spatial relationships, since virtual content appears properly anchored within the patient \cite{bajura1992merging, otsuki2015analysis, wang2017autostereoscopic, kalkofen2007interactive}. In general, focus-and-context techniques are considered intuitive and have received high usability ratings in user studies. However, excessive removal of occlusion layers can obscure the surrounding anatomy, potentially affecting task performance \cite{10108461}. 

Another class of techniques employs illustrative depth cues (e.g., color, shading, etc.) to convey spatial relationships without altering the geometry. Chromadepth rendering techniques, for instance, map depth to a color gradient (e.g., warm colors for nearer objects, cool colors for farther) to create a layered depth illusion. In neurosurgical AR, depth-encoded chromadepth coloring of cerebral vessels enabled surgeons to distinguish the depth and type of vessel during arteriovenous malformation resection \cite{kersten2012augmented}. Quantitative studies have shown that pseudo-chromadepth cues significantly improve depth discrimination in 3D medical images, outperforming even stereoscopic viewing for novice users \cite{kersten2013evaluation}. Unlike natural occlusion or shading, such illustrative cues require users to interpret an added visual code, but they can enhance spatial understanding without removing contextual structures.

A more direct approach to AR guidance is through explicit trajectory cues. These overlays depict the desired path or alignment of the instrument, often with real-time feedback on deviation. Many AR navigation systems in medicine draw virtual guide lines, target highlights, or entry point markers to explicitly direct instrument placement \cite{mewes2018concepts, acherki2025evaluation, zhang2024straighttrack}. In MRI-guided interventions, these approaches have been used to project colored alignment arrows and a dynamic depth gauge directly onto the patient’s body, turning green when the needle correctly matched both trajectory and depth \cite{mewes2018concepts}. Similar approaches have been used in spine procedures such as percutaneous vertebroplasty, where head-mounted displays projected preoperative trajectory plans in real time to guide needle insertion with improved accuracy \cite{abe2013novel}.

Despite a wide variety of AR visualization techniques that can be used for AR-guided percutaneous needle procedures, there is a lack of studies comparing their individual and combined benefits. Existing works often evaluate a single AR system against a conventional approach (e.g., AR versus external monitor guidance), demonstrating general benefits of AR as noted above. Some systems incorporate multiple visualization cues, such as combining trajectory guidance with focus and context overlays \cite{de2019augmented}, but do not explicitly assess the contribution of each component. This highlights a gap in comparative research, particularly in understanding how different visual cues affect precision, efficiency, and spatial awareness during AR-assisted percutaneous needle procedures.

In this work, we investigate how different visualization strategies affect user performance during AR-guided percutaneous needle procedures. These strategies, implemented in a unified AR system and evaluated individually and in combination, offer distinctive ways to support spatial understanding and help inform how to match visualization design to procedural needs.


\section{Methods} \label{sec:Methods}

To evaluate how different visual representations affect task performance during AR-guided percutaneous needle procedures, we implemented three visualization techniques: focus-and-context window (W), color-based proximity feedback (C), and trajectory validation (T). These designs were not arbitrarily chosen: they were developed through multiple rounds of discussion with medical experts who regularly perform procedures such as prostate biopsies. Clinicians emphasized that interpreting 2D images remains cognitively demanding for estimating depth and spatial relationships, even when using contemporary MRI–US fusion systems. Furthermore, they noted that many research AR systems rely on a naïve overlay of anatomy, which provides insufficient perceptual cues to enhance spatial understanding.

This feedback motivated our focus-and-context technique, which reveals internal anatomy through a localized cutaway to support depth perception. Clinicians also emphasized the need for real-time feedback on tool proximity to nearby anatomy, motivating our color-based proximity cue. For trajectory guidance, physicians noted that a direct cue from the tool to the target, especially one that conveys angular alignment, could help maintain the intended insertion path.


\subsection{Visualization techniques}
\subsubsection{Focus-and-Context Window (W)}

To resolve the depth ambiguity of internal targets while preserving global spatial context, we implemented a dual-component visibility system comprising a static surface window and a dynamic trajectory corridor. Surface Cutaway: A static, circular cutaway ($9\text{ cm}$ diameter) is rendered on the patient’s surface mesh. This window serves as a consistent visual entry point, featuring a smooth inward blue gradient at its boundary. This gradient is designed to provide subtle shading cues that reinforce the perception of surface curvature and depth, ensuring the "hole" feels anchored to the anatomy rather than floating in screen space. Dynamic Transparency Corridor: To facilitate the visualization of deep-seated internal structures, the system renders a semi-transparent cylindrical corridor centered on the planned needle path. The transparency within this volume is governed by a radial attenuation function that creates a gradual transition from the occluded surrounding tissue to the clear internal view. This effect is computed per voxel $x$ based on its radial distance $\rho(x)$ from the tool axis:
\begin{equation}
  \label{eq:window}
    \alpha_W(x) = \mathrm{lerp}(\alpha_{\text{min}}, \alpha_{\text{max}}, w(x)) \cdot \left(1 - M_{\text{cont}}(x)\right)
\end{equation}


\noindent where $\omega(x)$ is a weighting function derived from the corridor's inner and outer radii ($R_{\text{in}} = 16.5\text{ mm}$, $R_{\text{out}} = 21.5\text{ mm}$), and $\alpha_{\text{min}}, \alpha_{\text{max}}$ define the transparency range (set to $[0.13, 1.0]$). To maintain a localized spatial reference, a contact mask $M_{\text{cont}}(x)$ preserves a small, opaque disk ($3\text{ mm}$ radius) at the exact intersection of the tool and the skin surface. This ensures that the point of entry remains visible as a solid anchor, preventing the visual confusion caused by total transparency at the tool-tissue interface. The trajectory for the corridor is updated dynamically at each frame based on the tracked pose of the surgical tool, allowing the visualization to shift in real-time as the clinician navigates toward the target.





\subsubsection{Color-based Proximity Encoding (C)}

This technique encodes spatial proximity through color, without altering object geometry or transparency (\cref{fig:teaser}, middle column). It consists of two components: a surface heatmap and a target-specific blending cue.

\noindent\textbf{Surface Heatmap}: We implemented a custom shader that calculates the shortest radial distance $\rho(x)$ from each fragment $x$ on the patient's surface to the tool shaft. This distance is utilized to index a ChromaDepth-inspired spectral colormap, which leverages the chromostereopsis effect to enhance depth discrimination. The mapping assigns warmer colors (red) to nearer distances and cooler tones (green) to farther distances. The final fragment color $\mathbf{C}_{\text{surf}}(x)$ is computed by blending the anatomy's base color $\mathbf{c}_{\text{base}}$ with the colormap based on a maximum influence threshold $d_{\text{max}} = 0.1\text{m}$:
\begin{equation}
  \label{eq:color}
    \mathbf{C}_{\text{surf}}(x) = \mathrm{lerp}(\mathbf{c}_{\text{base}},\ \text{colormap}(\rho(x) / d_{\text{max}}),\ 1 - \rho(x) / d_{\text{max}})
\end{equation}

This ambient feedback allows clinicians to monitor the tool’s proximity to surrounding anatomy, such as the pelvic bone or bladder, providing a continuous spatial reference between the tool shaft and surrounding anatomical structures throughout the insertion process.\\

\noindent\textbf{Target Blending}: Complementing the surface heatmap, a goal-directed color cue is applied to the target (tumor) mesh. As the tool tip approaches the target center, its material color transitions from its original yellow ($\mathbf{c}_{\text{orig}}$) to a vibrant green ($\mathbf{c}_{\text{goal}}$). This blending uses a weight $t$ governed by a quadratic falloff to increase sensitivity during the final approach:
\begin{equation}
  \label{eq:color}
    \mathbf{C}_{\text{target}} = \mathrm{lerp}(\mathbf{c}_{\text{goal}},\ \mathbf{c}_{\text{orig}},\ t), \quad t = \left[ \min\left( \frac{d}{d_{\text{max}}},\ 1 \right) \right]^2
\end{equation}

where $d$ represents the Euclidean distance from the tool tip to the target center. By transitioning to green only upon close proximity, the system provides a clear, binary-like confirmation of alignment success while maintaining a continuous sense of distance through the quadratic transition.

\subsubsection{Explicit Trajectory Overlay (T)}

To provide explicit alignment guidance, this technique renders a dynamic, color-coded vector extending from the tool tip along its current orientation. This overlay supports angular awareness by projecting the intended insertion path directly into the anatomical space, particularly in regions where depth cues or surrounding structures may appear ambiguous. The visualization consists of a solid line (1 mm width, 20 cm length) whose color state is determined by real-time geometric validation against the target and critical anatomical regions. To ensure procedural safety, the system employs a hierarchical priority logic to define the line's color state:

\begin{itemize}
    \item \textbf{Critical Intersection (Red)}: This state maintains the highest priority. If the projected trajectory intersects any predefined danger regions, the line remains red regardless of target alignment. These intersections are computed via real-time raycasting against manually labeled critical volumes.
    \item \textbf{Target Alignment (Green)}: If no critical structures are intersected, the line turns green when the tool's forward vector achieves high-precision alignment with the target center. This is defined by a cosine similarity threshold $>0.999$, corresponding to an angular deviation of less than $2.5^\circ$. 
    \item \textbf{Suboptimal Orientation (Yellow)}: In the absence of both critical intersections and target alignment, the line is rendered in yellow, indicating a trajectory that is safe but requires further adjustment to meet the insertion criteria.
\end{itemize}

This hierarchical approach ensures that collision warnings with risk structures are prioritized over accuracy cues, providing clinicians with a clear, immediate assessment of trajectory validity during tool navigation.






\subsection{System Overview}
\label{system_overview}

\begin{figure}[tb]
  \centering
  \includegraphics[width=\linewidth]{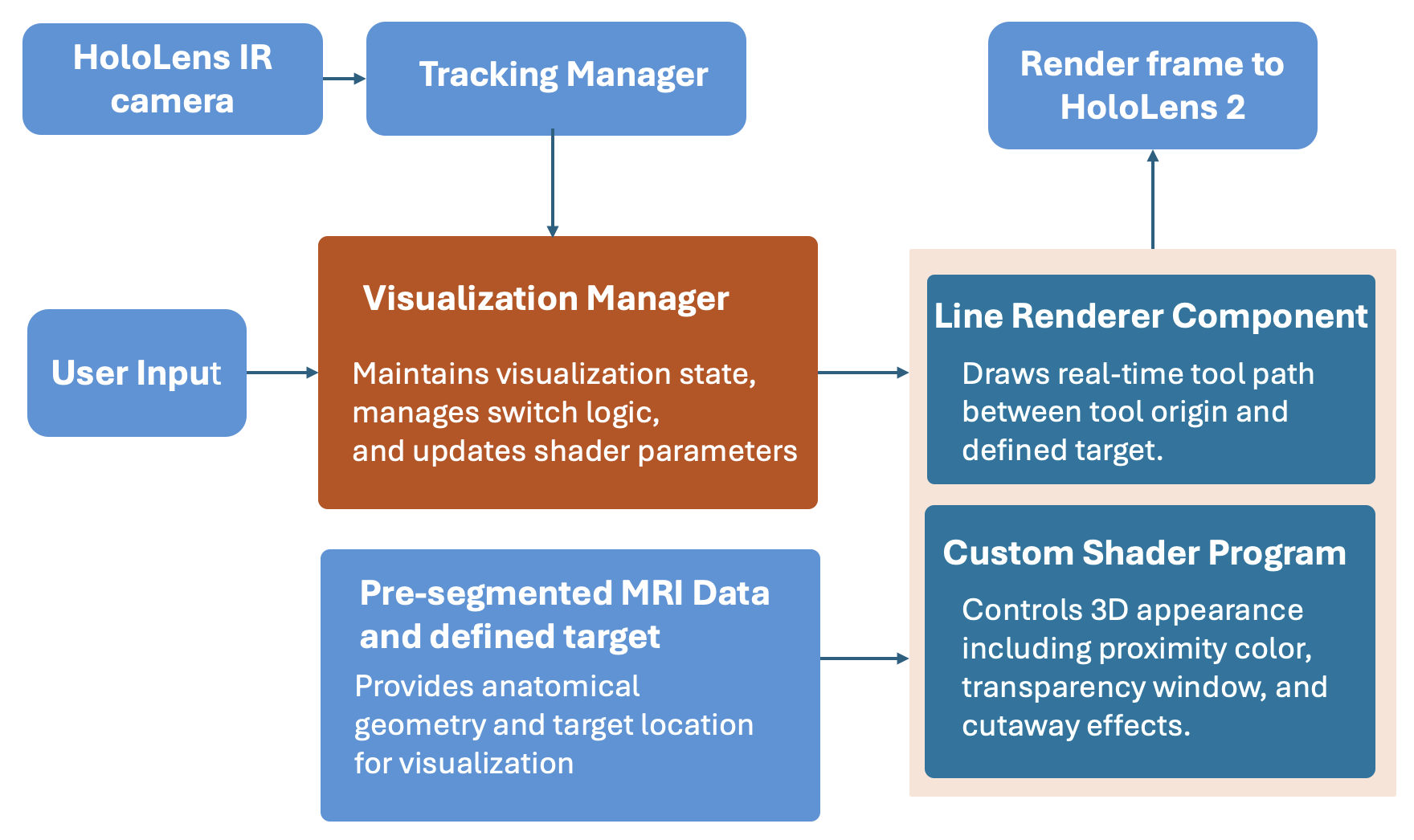}
  \caption{System architecture for visualization logic and rendering.  The diagram shows how tracking input, user input, and pre-segmented anatomical data are coordinated to control the trajectory overlay, proximity encoding, and focus-and-context visualization rendered on the HoloLens 2.
  }
  \label{fig:system_overview}
\end{figure}

The system was developed in Unity 2022.3 using MRTK3 and deployed on a HoloLens 2 head-mounted display. As shown in \cref{fig:system_overview}, the system consists of three main runtime inputs: (1) HoloLens IR camera data, (2) user input, and (3) pre-segmented MRI data with a defined target. 

A \textit{Tracking Manager} module processes camera-based tracking information, while the \textit{Visualization Manager} maintains the active visualization state, manages cue-switching logic, and updates shader parameters.
The \textit{Visualization Manager} then routes the active visualization state to two rendering components. The \textit{Line Renderer Component} draws the real-time tool path between the tool origin and the defined target, supporting the explicit trajectory overlay. The \textit{Custom Shader Program} controls the appearance of the anatomical models, including the color-based proximity encoding and the focus-and-context window. The resulting frame is rendered to the HoloLens 2 display.

\section{User Study: An Evaluation by Expertise }
\label{sec:user_study}
\begin{figure*}[h]
    \centering
    \includegraphics[width=1.0\linewidth]{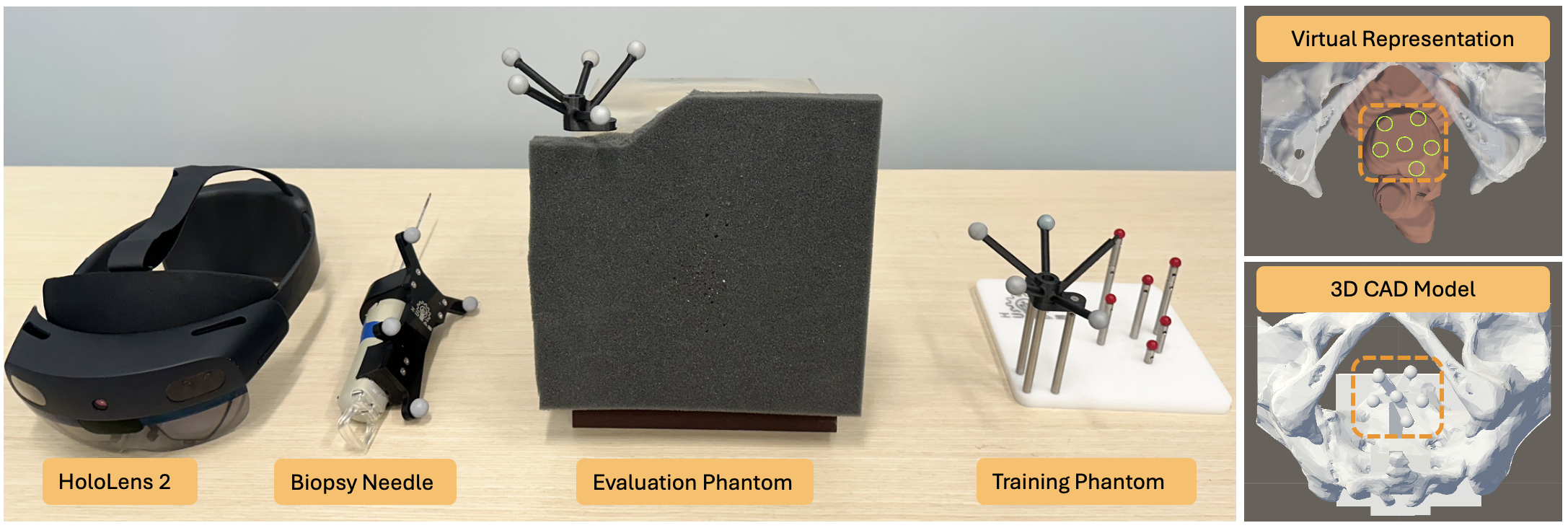}
    \caption{Experimental setup and anatomical modeling. Left: Hardware components used during the study, including the HoloLens 2 headset, tracked biopsy needle, and two custom phantoms. The evaluation phantom (center) contains a gel-embedded 3D-printed pelvis and internal target regions, while the training phantom (right) provides mechanical fixtures for initial practice. Right: The virtual representation used in AR overlays. A segmented 3D CAD model of the pelvis includes six spherical targets embedded within the prostate region (orange box), which align with physical targets inside the phantom.}
    \label{fig:experiment_setup}
\end{figure*}

To investigate how different visual augmentations affect task performance across levels of surgical experience, we conducted a stratified evaluation with two studies. Study I focused on a low-familiarity cohort of novices to assess baseline usability, while Study II examined the system's impact on a high-proficiency cohort of experienced practitioners. Both studies utilized a within-subjects experimental design involving a physical phantom and the Microsoft HoloLens 2. During the study, participants were asked to reach specific targets located inside the phantom using a biopsy needle for three rounds (\cref{fig:experiment_setup}). Each round included the use of all possible W/C/T cue combinations (\cref{fig:teaser}). A total of 24 trials were performed per participant. The study was conducted in a private office, where external distractions were reduced. The study protocol was approved by the Institutional Review Board (IRB) of Johns Hopkins University (HIRB00022863).

\subsection{Participants}

\noindent\textbf{Study I: Evaluation with Novices.}
In Study I, we recruited nineteen participants (\(N=19\)) with no prior or very limited experience in surgical needle procedures. This cohort included non-medical volunteers and medical students who had only observed such procedures. Participants’ ages ranged from 22 to 58 years (\(M=29.2, SD=8.3\)), including 13 male and 6 female participants. All participants reported normal or corrected-to-normal vision and no upper limb impairments.
\\ \\
\noindent\textbf{Study II: Evaluation with Clinically Trained Experts.}
In Study II, we recruited seven clinically trained participants (\(N=7\), referred to as experts in the remainder of the paper) with prior experience in percutaneous needle procedures and familiarity with the prostate biopsy workflow. This cohort consisted of two attending surgeons, four clinical fellows, and one senior medical student who had performed multiple prostate biopsies under clinical supervision. Participants’ ages ranged from 24 to 37 years (\(M=31.7, SD=3.9\)), including 6 male and 1 female participant.

\subsection{Experimental Setup}
We designed a physical phantom replicating the anatomy observed when performing prostate biopsy,  containing 3D-printed pelvic structures and six $4\text{ mm}$ radius spherical targets submerged in a gelatin-based tissue simulant (\cref{fig:experiment_setup}). The visualization techniques described in \cref{sec:Methods} were implemented using Unity 3D and presented to the study participants using the Microsoft HoloLens 2. The physical phantom and a biopsy needle were equipped with passive IR markers and tracked using the built-in sensors of the Microsoft HoloLens 2 \cite{10021890}.

\subsection{System Validation}
\label{sec:system_validation}

Before conducting the user study, we performed a validation stage to ensure that the printed phantom's geometry and the tracking accuracy of the system would contribute to the acquisition of reliable data.

To identify the physical location of the tool tip in the coordinate frame of the optical tracker, we conducted a pre-validation test using a RANSAC-based pivot calibration, yielding a tooltip localization error of 0.20 mm (RMSE). To evaluate targeting accuracy under controlled yet representative conditions, we used the evaluation phantom with the top gel layer removed to eliminate resistance and deformation that might influence needle shape. One of our researchers manually aligned the tool tip with the center of a predefined spherical target 78 times.

System error was defined as the Euclidean distance between the calibrated tip and the target center. To reduce transient hand motion noise and better reflect stabilized tool placement, we collected and averaged the final three frames after user-confirmed completion. Across 78 repeated trials, the mean targeting error was 3.70\,mm $\pm$ 1.37\,mm (mean$\pm$SD; SE=0.16\,mm). Relative to the theoretical lower bound of 4 mm (target radius), the signed bias was $-0.30\mathrm{mm}$ (95\% CI $[-0.61, 0.01]\,\mathrm{mm}$), with a residual RMSE of $1.40\,\mathrm{mm}$.

\subsection{Study Design}
\label{section:experimental_design}
We used a within-subjects 2×2×2 factorial design corresponding to the three techniques described in \cref{sec:Methods}. Each factor had two levels (off/on), yielding a total of eight experimental conditions:
\begin{itemize}
    \item Context Window (Window: off/on): when on, a localized cutaway around the needle tip reveals occluded anatomy; when off, the organ surface remains intact.
    \item Color Feedback (Color: off/on): when on, the color of the target's surface is modulated as a function of the distance between its location and the needle tip; when off, default shading is used.
    \item Trajectory Visualization (Trajectory: off/on): when on, the ideal path and the current needle trajectory are rendered with semantic color coding; when off, no trajectory cues are shown.
\end{itemize}

The eight possible combinations of visual cues W/C/T, including a baseline representation showing a naive representation of the virtual target (i.e., W/C/T = \texttt{000}), were presented once per round in a randomized order. Participants were asked to use the needle to reach the target using each of these combinations for a total of three rounds, resulting in the completion of 24 trials per participant (8 conditions $\times$ 3 rounds). The objects' geometry and viewing conditions were held constant across trials. The order of appearance of the visualization combinations and the target was randomized every round as a means to mitigate sequence and learning effects (\cref{fig:teaser}). Participants were not informed which visualization(s) were active in a given trial.

\subsection{Procedure}
Participants were briefed on the study objectives and provided with an overview of the system and interface. A brief introductory walkthrough using slideshows illustrated the visualization components (color feedback, context window, trajectory line) without disclosing the experimental conditions. Participants then received detailed instructions on how to manipulate the tool and execute the task.

\subsubsection{Pre-task Familiarization}
Participants were given the opportunity to familiarize themselves with the system and tools by completing two needle insertion trials using the baseline visualization on two distinct phantoms designed for different purposes (See \cref{fig:experiment_setup}):

\begin{enumerate}
    \item \textit{Training phantom (visible target)}: Participants practiced placing the needle's tool tip on visually indicated targets to experience the virtual guidance and hand–eye mapping.
    \item \textit{Evaluation phantom (invisible target):} With the physical target hidden in the experimental phantom, a virtual line from the target toward the intended entry path was rendered. Participants were then asked to insert the tool following this trajectory to experience resistance and guided motion.
\end{enumerate}
The data generated from these trials were logged (as described in \cref{system_overview}) for system verification and task execution monitoring, but were not used for data analysis.

\subsubsection{Main Experiment}
Participants were tasked with guiding the tool tip to an internal phantom target using the visualization(s) active for that trial. Conditions were assigned and ordered according to the within-subjects randomization procedure described in \cref{section:experimental_design}. Participants were not informed which visualization(s) were enabled.

A trial began when the participant initiated needle insertion and ended upon their verbal confirmation that the tool had reached the target. Throughout the trial, the system continuously logged the pose of the needle's tool tip and recorded objective measures, including final target error and task time. During the first round of trials, participants completed the NASA-TLX questionnaire. In Rounds 2 and 3, they reported the perceived ease of use and usefulness of the visualization using two separate single-item questions on a 7-point Likert scale. At the end of the session, participants completed the System Usability Scale (SUS) questionnaire to assess overall system usability. Finally, we conducted short interviews to collect open-ended feedback regarding their experience.

\subsection{Experimental Variables}
\label{sec:metrics}

Three objective metrics were collected and computed using per–trial logs of the tool tip pose collected at 10 Hz and saved locally as comma-separated values files in the HoloLens 2. In addition, we collected subjective metrics via questionnaires to assess the participants' perceived mental load, task difficulty, and system usability. 

\subsubsection{Objective metrics}
\label{sec:metrics_objective}
\begin{itemize}
    \item Final center error. Defined as the Euclidean distance, in millimeters, between the final tool-tip position at trial completion and the target center in world coordinates. Because the physical target is a 4 mm radius sphere embedded in the phantom, this metric has a theoretical floor of 4 mm (the tool tip cannot physically reach the center of the 3D printed target). 
    \item Time to completion. Elapsed time, in seconds, from the start of the programmatic trial to the participant's confirmation of completion.
    \item Retreat events. A retreat event was considered as any intentional and contiguous segment during which the distance to the target center monotonically increases (i.e., moves in the opposite direction), and the cumulative along-path displacement within that segment exceeds 30 mm. Each qualifying segment exceeding this distance counted as a single instance.
\end{itemize}

\subsubsection{Subjective metrics}
\label{sec:metrics_subjective}
\begin{itemize}
    \item NASA-TLX. We used the NASA task load index questionnaire (TLX) \cite{hart1988development} using a 21-point rating scale (0–20) to assess and derive assumptions of the task load, and analyzed its individual sub-scales (Raw-TLX) \cite{hart2006nasa} to avoid introducing additional sources of measurement errors \cite{bustamante2008measurement}. This information was collected after each trial during Round 1.
    \item Single-item ratings. During Rounds 2 and 3, we collected the participants' perceived task difficulty after each trial using the Single Ease Question (SEQ) \cite{wetzlinger2014comparing}. This questionnaire (“Overall, this trial was: ") consists of a 7-point Likert scale ranging from 1 (Very easy) to 7 (Very complicated). In addition, we presented participants with a similar questionnaire asking to rate the usefulness of the visualization (“The visualization(s) were useful for this trial: ”) using a 7-point Likert scale ranging from 1 (Not Helpful) to 7 (Very Helpful).
\end{itemize}

\section{Data Analysis}
\subsection{Hypotheses and analysis overview}
Effective visualization should help users complete tasks more accurately, efficiently, and with reduced cognitive effort. In this study, we define performance improvement as achieving (i) lower final center error, (ii) shorter time to completion, and (iii) fewer tool retreats. To evaluate subjective experience, we also analyzed subscales from the NASA–TLX, the Single Ease Question (SEQ), and the single-item usefulness rating (Likert scale) described in \cref{sec:metrics}. These experimental variables served to test the following hypotheses:

\begin{enumerate}\setlength\itemsep{2pt}
  \item[\textbf{H1.}] Any technique incorporating one or more cues will outperform the naive, baseline visualization, in performance.
  \item[\textbf{H2.}] Cue combinations will yield benefits beyond single cues, as they convey different information (occlusion/visibility, distance-to-target, trajectory/direction). 
  \item[\textbf{H3.}] Relative to single cues and baseline, cue combinations will lead to improvements in subjective metrics.
\end{enumerate}

\subsection{Statistical modeling} 
\label{sec:statistical modeling}

We analyzed objective (error, time, retreats) and subjective (NASA-TLX, SEQ) metrics using Linear Mixed Models (LMMs) \cite{rasch2025ar}. Window, Color, and Trajectory were modeled as fixed effects in a full-factorial $W$$\times$$C$$\times$$T$ design. To account for learning effects and task difficulty, Round (1–3) and Target ID ($T_n, n \in [1,6]$) were included as fixed-effect covariates. Participant ID was specified as a random effect to control for repeated measurements. To ensure model stability given the small expert sample ($N=7$), $T_n$ was treated as a fixed rather than random effect. LMMs were estimated via Restricted Maximum Likelihood (REML) with Satterthwaite degrees of freedom for Type-III $F$-tests. For retreat counts, we employed a negative-binomial Generalized Linear Mixed Model (GLMM) with Type-III Wald $\chi^2$ tests. Post-hoc analyses and simple-effects contrasts were performed using R's \textit{emmeans}, with numeric covariates held at their means. To evaluate expertise-related differences, pooled models included Cohort (expert, novice) and its interactions with cues as fixed effects. Multiple comparisons were corrected using Holm’s method for pairwise contrasts and Dunnett’s test for baseline comparisons. Estimates are reported as back-transformed Estimated Marginal Means (EMMs) and ratios (GMR/IRR) with 95\% CIs. Model assumptions were verified via residual Q–Q plots.

\begin{figure}
    \centering
    \includegraphics[width=0.5\textwidth]{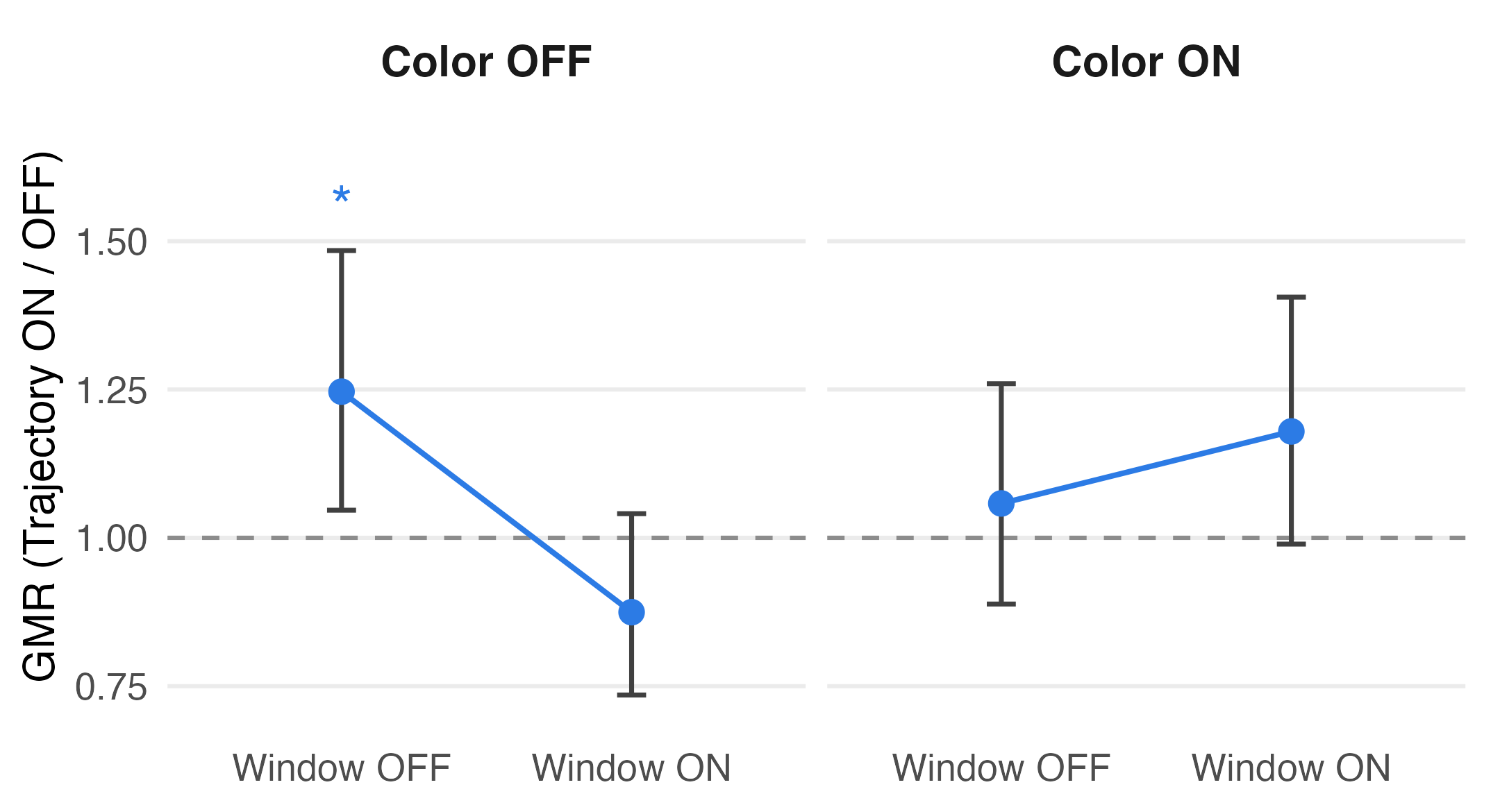}
    \caption{Trajectory simple effects on task duration in non-experts. Points and 95\% confidence intervals show GMRs comparing Trajectory ON to Trajectory OFF within each Window × Color condition. GMR > 1 indicates longer task duration with Trajectory. The dashed line indicates no difference. Asterisks denote significant simple effects.}
    \label{fig:traj_effect}
\end{figure}

\section{Results}
\begin{figure*}[t]
    \centering
    \includegraphics[width=1.0\textwidth]{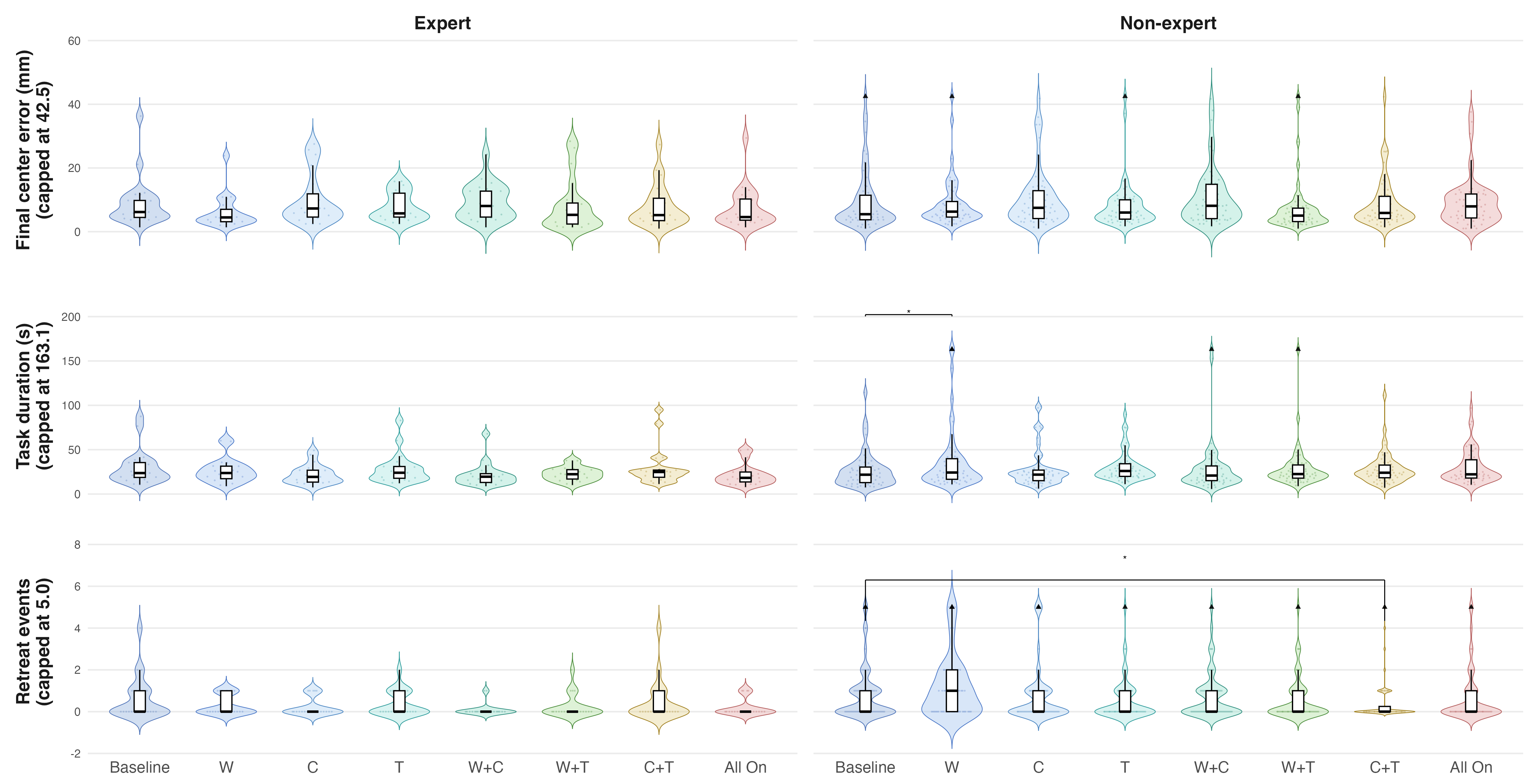}
    \caption{Objective metrics by visualization condition and expertise. Violin plots show the distributions of final center error, task duration, and retreat events across the eight cue configurations (Baseline, W, C, T, W+C, W+T, C+T, All On). Rows correspond to objective metrics, and columns correspond to experts (Study II) and non-experts (Study I). Boxes indicate medians and interquartile ranges; points show individual trials. For readability, values above the cap are shown at the cap noted on each row label. Brackets denote significant Dunnett-adjusted comparisons relative to the baseline condition within each cohort.}
    \label{fig:violin_graph}
\end{figure*}

To assess the impact of visualization techniques on task performance, we first analyzed the distributions of objective metrics across conditions (\cref{fig:violin_graph}). While this descriptive view provides intuition about potential differences, formal inference was conducted through Linear Mixed Models (LMMs), which test the fixed effects of Window, Color, and Trajectory, their interactions, and the covariates described in \cref{sec:statistical modeling}. Both studies used the same $W\times C\times T$ factorial design and the same mixed-effects modeling approach. We report results separately for non-experts (Study~I) and experts (Study~II), followed by a pooled analysis to assess whether cue effects differed by cohort. Across these analyses, we focus on the main patterns most relevant to the hypotheses. Detailed omnibus tests and targeted follow-up analyses are reported in the supplemental material.

\subsection{Study I: Performance with Novices}
\noindent\textbf{Accuracy (Final Center Error).}
Accuracy (Final Center Error). For non-experts, Trajectory reduced final center error, while Color increased final center error. Trajectory reduced final center error by approximately 13\% (GMR = 0.870, 95\% CI [0.788, 0.960], $p=.0056$), while Color increased final center error by approximately 14\% (GMR = 1.139, 95\% CI [1.032, 1.257], $p=.0096$). Window did not significantly affect final center error.\\

\noindent\textbf{Time to Completion.}
Completion time decreased across rounds. The effect of Trajectory depended on whether Window and Color were also present (\cref{fig:traj_effect}). When Window and Color were both absent, adding Trajectory increased task duration by approximately 25\% (GMR = 1.246, 95\% CI [1.047, 1.484], $p=.014$). When Window or Color was enabled, Trajectory did not significantly change completion time. When each visualization condition was compared with the baseline condition, the Window-only condition was significantly slower (GMR = 1.281, 95\% CI [1.014, 1.618], $p_{\text{Dunnett}}=.032$).\\


\noindent\textbf{Retreat Events.}
For non-experts, Trajectory significantly reduced retreat events by approximately 41\% (IRR = 0.594, 95\% CI [0.445, 0.793], $p<.001$). Only Color+Trajectory significantly reduced retreat events when compared with the baseline condition (IRR = 0.434, 95\% CI [0.193, 0.974], $p_{\text{Dunnett}}=.040$), corresponding to an estimated 57\% reduction. The remaining omnibus effects, including the marginal Color effect and the Round effect, are reported in the supplemental material.\\


\noindent\textbf{Subjective Experience (NASA--TLX and SEQ).}
For workload (NASA--TLX, Round 1), none of the cue-related main effects or interactions were statistically significant after within-term FDR correction (all $q \ge .120$), suggesting no detectable reduction in perceived workload attributable to cueing. In contrast, subjective experience (SEQ, Rounds 2--3) was selectively influenced by cue presentation. Color significantly improved SEQ Ease ($F_{1,260.00}=10.11, q=.002$) and strongly increased SEQ Helpfulness ($F_{1,260.02}=55.47, q < .001$). Trajectory also significantly enhanced Helpfulness ($F_{1,260.03}=25.23, q < .001$). However, Window showed no significant effect on either SEQ Ease or Helpfulness (all $q \ge .254$).

\subsection{Study II: Performance with Experts}
\noindent\textbf{Accuracy (Final Center Error).}
None of the visual cues significantly changed the final center error in experts (Window: $p=.100$; Color: $p=.189$; Trajectory: $p=.262$). All GMR effect-size estimates had 95\% CIs encompassing 1.\\


\noindent\textbf{Time to Completion.}
Experts exhibited strong learning effects, with completion times decreasing significantly across rounds ($p < .001$). Both Window and Color significantly reduced completion time, yielding comparable improvements of approximately 12–13\% (Window: $p=.023, \text{GMR}=0.877$; Color: $p=.017, \text{GMR}=0.871$). In contrast, Trajectory had no detectable impact ($p=.662$). No significant interactions were observed (all $p \ge .155$).\\

\noindent\textbf{Retreat Events.}
Window was the only cue that significantly reduced retreat events in experts ($\text{IRR}=0.475$, 95\% CI [0.247, 0.911], $p=.025$), corresponding to an estimated 52\% reduction. Color and Trajectory did not significantly affect retreat events.\\

\noindent\textbf{Subjective Experience (NASA--TLX and SEQ).}
For SEQ Ease (Rounds 2--3), Trajectory was the only cue to significantly improve perceived ease after FDR correction ($F_{1,94.02}=5.45, q=.043$), while Window and Color had no detectable effect ($q \ge .183$). For SEQ Helpfulness, both Trajectory ($F_{1,94.02}=26.59, q<.001$) and Color ($F_{1,94.05}=6.10, q=.031$) significantly increased ratings, whereas Window did not ($q=.215$). Regarding workload (NASA--TLX, Round 1), most subscales showed no cue effects; however, Window significantly improved perceived performance on the TLX Performance subscale ($F_{1,39.19}=10.09, q=.017$).

\begin{figure}[h]
    \centering
    \includegraphics[width=0.5\textwidth]{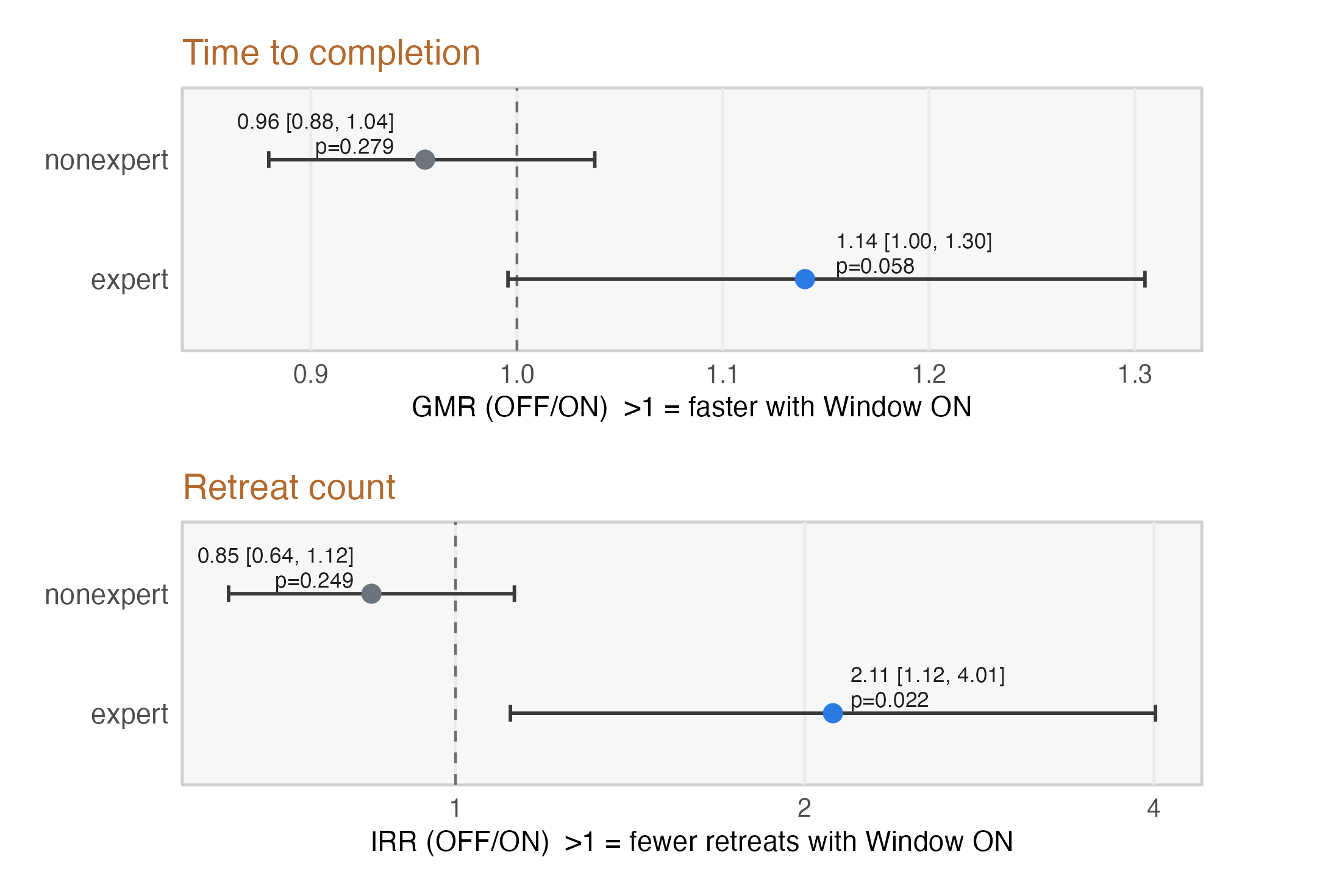}
    \caption{Points and horizontal bars represent model-estimated ratios (Time: GMR; Retreat: IRR) with 95\% CIs, comparing Window OFF vs.\ ON for experts and non-experts. The vertical dashed line at 1 denotes the null effect; ratios $>1$ indicate improved performance with Window ON (i.e., faster completion or reduced retreats). Annotated values provide exact ratio estimates and $p$-values.}
    \label{fig:windwo_effect}
\end{figure}

\subsection{Generalization across Cohorts}
To assess whether cue effects differed by expertise, we fit pooled models including Cohort (expert vs.\ non-expert) and cue$\times$Cohort interactions for each objective metric. For accuracy, there was no evidence that cue effects varied by cohort (all cue$\times$Cohort interaction terms: $p \ge .245$). In contrast, retreat behavior showed cohort-dependent effects (\cref{fig:windwo_effect}): the pooled retreat model indicated significant Window$\times$Cohort interaction ($p=.010$), suggesting that the impact of Window on retreat events differed between experts and non-experts. Completion time also showed heterogeneity across cohorts, with a significant Window$\times$Cohort interaction in the pooled time model ($p=.029$). 

\subsection{System Usability Scale (SUS)}
Across participants with complete SUS responses (n=25), the system achieved a mean SUS of $72.8$ (SD=$13.8$), which is above the commonly used “average usability” benchmark of 68. SUS scores were comparable between experts (M=$71.1$) and non-experts (M=$73.5$), with overlapping confidence intervals, suggesting broadly similar perceived usability across cohorts. The observed spread in scores (37.5–95.0) indicates that while many users found the system highly usable, a subset experienced notable friction, motivating further refinement and robustness improvements.



\section{Discussion}

\textbf{H1.} Overall, H1 received mixed support across the objective metrics. Trajectory cueing benefited novices on accuracy and retreats (e.g., improved final center error), but other cues introduced costs in specific conditions (e.g., Window-only increased completion time), and no cue uniformly improved performance across metrics and expertise levels.
\\ \\
\noindent\textbf{H2.} Evidence for additive or synergistic benefits from cue combinations was limited. Cue effects were often non-additive (including a significant three-way interaction in completion time), and only specific combinations improved specific metrics, rather than yielding broad, consistent gains across outcomes.
\\ \\
\noindent\textbf{H3.} Subjective data suggested perceived benefits of individual cues (Color and Trajectory in novices, and Trajectory helpfulness in experts), but provided limited evidence that multi-cue configurations systematically improved subjective outcomes beyond single cues.

\subsection{The Expertise Reversal Effect in Surgical AR}
The results suggest that cue effects depended on expertise and can be discussed in relation to the Expertise Reversal Effect (ERE) in cognitive load theory~\cite{kalyuga2009expertise}. ERE describes how instructional supports that benefit novices can be less useful for more experienced users.
Nevertheless, the observed between-cohort pattern suggests that
trajectory guidance produced measurable benefits for novices, whereas any benefit for experts, if present, was not comparably pronounced in our data, consistent with the “diminishing returns” pattern emphasized by ERE~\cite{kalyuga2009expertise}.

The focus-and-context cue showed a complementary pattern. For novices, the window did not improve accuracy and increased completion time relative to baseline when presented alone (GMR$=1.281$, $p_{\text{Dunnett}}=.032$). This suggests that interpreting occlusion relationships and depth ordering may impose integration costs that exceed the potential benefits early in the learning process. Such a finding aligns with the account that search and state-updating demands can occupy limited working memory resources, leaving fewer resources available for constructing a coherent spatial representation~\cite{sweller1988cognitive}. For experts, in contrast, the window reduced completion time (GMR$=0.877$, $p=.0231$) and retreat events (IRR$=0.475$, $p=.025$). One plausible explanation is that experts primarily used the window as a quick verification cue to confirm alignment, rather than as information that required extensive additional interpretation.

\subsection{The Color Paradox}
The novice results reveal a dissociation between subjective preference and objective performance for the color cue. Although non-experts rated the surface heatmap as easier and more helpful (SEQ Ease: $p=.0010$; SEQ Helpfulness: $p=5.76\times10^{-13}$), enabling color was associated with higher final center error ($p=.0096$). This subjective sense of utility was reinforced by qualitative observations, as some participants noted in post task interviews that the green color change provided a reassuring signal that alignment was complete. A plausible explanation is that the continuously varying “greener-is-better” gradient acted as a highly salient, locally reinforcing signal, biasing attention toward near-surface proximity and away from subtler error signals needed for fine-grained motor correction, consistent with attentional tunneling and related inattentional effects reported for augmented displays and surgical AR \cite{dixon2013surgeons,wickens2009attentional}. 
In this task, novices already needed substantial visual-spatial resources to interpret depth, occlusion, and tool motion. Adding a vivid, spatially dense heatmap may therefore have increased competition within the same visual channel, elevating visual workload and reducing attentional flexibility. Together, these mechanisms could make the cue feel supportive while encouraging a control strategy that over-optimizes local proximity feedback rather than the global endpoint criterion, resulting in worse final accuracy. This pattern is unlikely to reflect a speed-accuracy tradeoff because color did not reliably reduce completion time in novices ($p=.260$). In experts, we did not detect a reliable accuracy effect of color ($p=.189$), although the small expert cohort limits precision.

\subsection{Feedforward Trajectories and Perceived Utility}
While the trajectory cue did not yield measurable changes in expert accuracy or completion time, experts nevertheless reported high perceived utility (e.g., SEQ Helpfulness: $p<.001$). This pattern may reflect the functioning of the trajectory primarily as \emph{a feedforward}: by making the intended action and alignment explicit, the cue could ease the translation of an intention into an executable motor command, even when the performance of the endpoint is already near the ceiling \cite{vermeulen2013crossing}. In other words, the trajectory may support action selection and alignment by clarifying “how to act next,” which can increase perceived helpfulness without necessarily producing detectable changes in objective outcomes. This interpretation is especially plausible for experts, whose baseline performance may already be stable and for whom our outcome measures may be less sensitive to changes in planning and coordination demands.

\subsection{Design Implications}
Although our findings are specific to the visualizations tested in a controlled setting, they highlight several considerations that may guide the design of augmented reality systems for procedural tasks that follow the mechanical action performed in this study. Rather than proposing fixed design rules, we share observations that could inform how to balance visual cues, reinforce user understanding, or increase system usability.
\\ \\
\textit{Expertise-adaptive cueing.} The two cohorts exhibited distinct performance patterns, suggesting that cue presentation should reflect user proficiency. For novices, trajectory guidance improved accuracy and reduced retreats, indicating value as an early-stage support. In contrast, the focus-and-context window imposed a time cost for novices when presented alone, while benefiting experts in efficiency and retreats. This pattern motivates interfaces that introduce complex contextual information progressively, after users have developed more stable task-specific mental models and can integrate occlusion and depth cues at lower cost.
\\ \\
\textit{Managing feedback salience.} The dissociation between novice preference and accuracy for the color heatmap suggests caution when using highly salient proximity feedback. Although novices perceived the heatmap as easier and more helpful, it was associated with higher final error, consistent with the possibility that attention is drawn toward locally reinforcing proximity signals rather than the global centering objective. Designers may therefore consider reducing the heatmap’s visual dominance during critical alignment phases (e.g., lower saturation, reduced spatial density, smaller footprint, or on-demand display), and prioritizing encodings that more directly reflect the task criterion (e.g., predicted centering error) rather than proximity alone.
\\ \\
\textit{Speed and Precision trade-offs.} Visual aids may not improve time and accuracy simultaneously. For novices, trajectory guidance was associated with higher precision but also an $\sim$25\% increase in completion time when presented without other cues. This suggests that developers should decide whether the primary goal is speed, precision, or a controlled balance, and tailor guidance accordingly. For speed-critical settings, lighter-weight or context-triggered trajectory presentation may be preferable, whereas precision-critical tasks may justify more persistent guidance despite time costs.
\\ \\
\textit{Process-oriented metrics and usability.} Process measures such as retreat counts provided a complementary view of cue utility beyond final error, capturing differences in procedural flow that were not always reflected in endpoint accuracy. In both cohorts, high helpfulness ratings for trajectory (and for color in some cases) indicate perceived support even when objective gains were limited. At the same time, Round~1 TLX did not show clear cue-related workload differences for novices, suggesting that perceived demand remained high early on. These findings motivate evaluating AR guidance with a combination of final outcome metrics and process-oriented measures (e.g., retreat events), alongside subjective ratings that capture perceived coordination and cognitive friction during execution.

\section{Limitation and Future Work}
Our study was conducted on a rigid phantom in a controlled environment, which allowed us to isolate visualization effects but limits how well the results may carry over to real-world procedures. In the operating room, clinicians face additional challenges such as soft-tissue deformation, organ motion, variable tactile feedback, patient response, imaging artifacts or uncertainty and time pressure. These factors influence behavior—for example, the retreat behavior we observed may occur less often during live procedures, where retries are discouraged. Since our setup did not include tissue dynamics or imaging artifacts, future work should test the system under more realistic conditions, such as cadaver studies or clinical trials.

We compared different visualization techniques by having participants perform a hands-on needle guidance task, allowing us to directly evaluate their impact on task performance. However, the specific visual designs we implemented may not represent the most effective versions of each technique. Future studies should explore alternative design variations to better understand which visual features contribute most to usability, accuracy, and efficiency. In addition, while we explain the color paradox via attentional tunneling, we did not directly measure attention allocation (e.g., eye tracking) or manipulate heatmap salience to causally test this mechanism. Future work will vary salience parameters (saturation/footprint/discretization) and add attention measures to validate whether attentional capture mediates the observed accuracy degradation.

This study compared separate non-expert and expert cohorts in a single session. Therefore, the results do not show how the same users would adapt to the cues over training. Future work should use longitudinal studies to examine how cue reliance, trust, task duration, speed--precision trade-offs, and workload change with repeated AR-guided needle-alignment practice. These studies could also test whether cue configurations should change as users gain experience.

Finally, our sample size was modest (\textit{N}=26; 7 experts and 19 non-experts), and the participant pool included a higher proportion of male participants, particularly among experts. This distribution reflects the demographics of our local recruitment channels and the gender composition often observed in urology-related clinical contexts. Future work with larger and more diverse samples will be important to assess population-level effects and improve generalizability across users and environments.

\section{Conclusion}

In this work, we evaluated three augmented reality visual cues for percutaneous guidance using a within-subjects factorial design. Our results show that the effects of these cues were not uniform, but depended on both user expertise and cue configuration. For novices, trajectory guidance improved targeting accuracy and reduced retreat behavior, but could also increase completion time, indicating a speed–precision trade-off. For experts, the focus-and-context window was associated with faster completion and fewer retreats, while color feedback was associated with faster performance. Across cohorts, combining cues did not produce consistent additive benefits, and subjective preference did not always align with objective performance. In particular, color cues were often perceived as helpful even when their effects on accuracy were not consistently positive. Overall, these findings suggest that AR guidance for needle-based procedures should be designed as expertise-aware and context-aware, rather than by simply adding more visual information. Future work should validate these effects in more realistic procedural settings and with larger expert cohorts, and further examine how cue design and visual salience influence both procedural flow and endpoint performance.

\acknowledgments{%
  This work was supported by the Graduate Partnership Program (GPP) of the National Institutes of Health (NIH) and the National Institute of Biomedical Imaging and Bioengineering (NIBIB).}%

\bibliographystyle{abbrv-doi}

\bibliography{template}
\end{document}